\documentclass[runningheads]{llncs}

\usepackage{graphicx,xcolor}

\usepackage{amsmath,latexsym}
\usepackage[all, warning]{onlyamsmath}

\usepackage[linesnumbered,plain,vlined,figure]{algorithm2e}

\usepackage{booktabs,multirow}
\usepackage{cite,url}

\newcommand{\Sec}[1]{Sect.~#1}
\newcommand{\Fig}[1]{Fig.~#1}

\newcommand{\Eq}[1]{Eq.~(#1)}

\DeclareMathOperator{\Table}{\mathcal{T}}
\DeclareMathOperator{\Column}{\mathit{A}}
\DeclareMathOperator{\Columns}{\mathbf{A}}
\DeclareMathOperator{\Value}{\mathit{a}}

\DeclareMathOperator{\ZValue}{\mathit{z}}
\DeclareMathOperator{\Ranges}{\mathit{R}}

\begin{document}
\title{Z-ordered Range Refinement\\for Multi-dimensional Range Queries}
%
\author{
  Kento Sugiura
  \and Yoshiharu Ishikawa
}
\authorrunning{Kento Sugiura and Yoshiharu Ishikawa}
\institute{
  Graduate School of Informatics, Nagoya University
}
\maketitle

\begin{abstract}
  The z-order curve is a space-filling curve and is now attracting  the interest of developers because of its simple and useful features.
  In the case of key-value stores, because the z-order curve achieves multi-dimensional range queries in one-dimensional z-ordered space, its use has been proposed for both academic and industrial purposes.
  However, z-ordered range queries suffer from wasteful query regions due to the properties of the z-order curve.
  Although previous studies have proposed refining z-ordered ranges, doing so is computationally expensive.
  In this paper, we propose \emph{z-ordered range refinement} based on \emph{jump-in/out} algorithms,  and then we approximate  z-ordered query regions to achieve efficient range refinement.
  Because the proposed method is lightweight and pluggable, it can be applied to various databases.
  We implemented our approach using PL/pgSQL in PostgreSQL and evaluated the performance of range refinement and multi-dimensional range queries.
  The experimental results demonstrate the effectiveness and efficiency of the proposed method.

\end{abstract}

\section{Introduction}\label{sec:introduction}

With the development of hardware in recent years, industrial and academic researchers have developed various new databases.
The latest CPUs have 50 or more physical cores, and a single server can be equipped with terabytes of memory; this new hardware has brought about changes in architecture, leading to the development of new database systems such as in-memory relational database management systems~\cite{sosp:Tu2013,sigmod:Neumann2015} and key-value stores~\cite{url:rocksdb}.
However, a problem associated with developing these new databases is the implementation of complex data types and indices.
In particular, because key-value stores usually have restricted features, certain techniques must be used to implement the desired applications.

For multi-dimensional data types, the z-order curve~\cite{tech:Morton1966} is now attracting the interest of developers because of its simple and useful features.
The z-order curve is a space-filling curve that allows multi-dimensional data to be processed in one-dimensional z-ordered space.
That is, multi-dimensional data can be used in key-value stores by using z-ordered keys.
Given that it is also easy to convert multi-dimensional range queries into z-ordered ones, several previous studies have used the z-order curve in their implementations~\cite{vldb:Armbrust2020,ieice:Le2018,dpdb:nishimura2013}.

\begin{figure}[t]
  \includegraphics{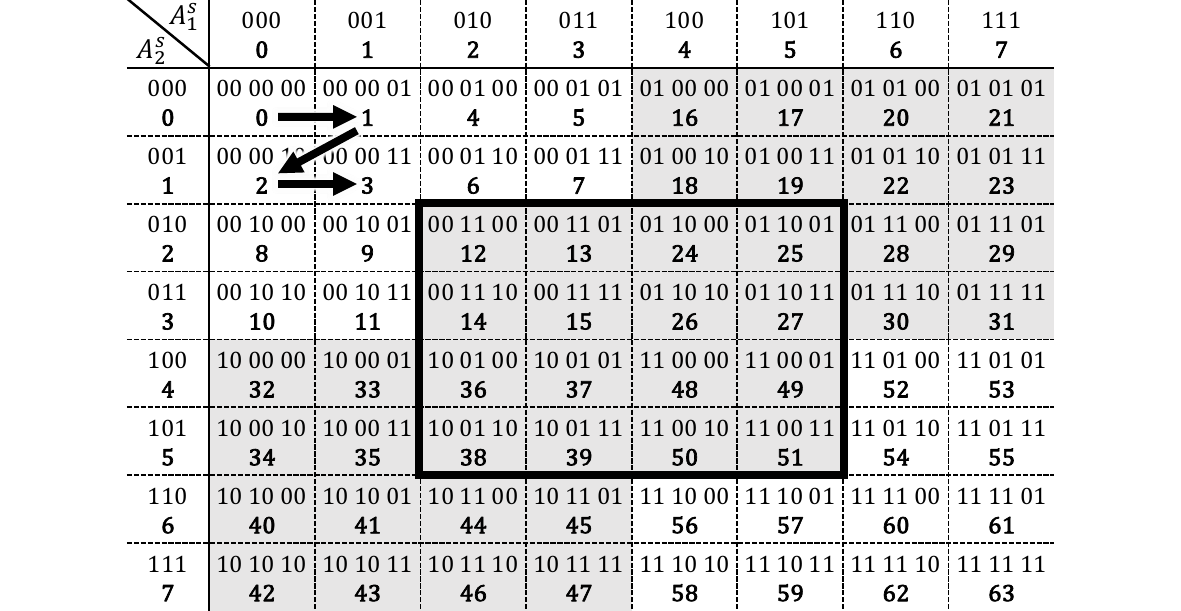}
  \caption{Two-dimensional z-order curve space and a query region}
  \label{fig:z-curve}
\end{figure}

The problem with z-ordered range queries is wasteful query regions due to the properties of the z-order curve.
An example of wasteful regions in a z-ordered range is shown in \Fig{\ref{fig:z-curve}}, where the solid rectangle shows the original multi-dimensional query region, but its z-ordered query region (gray cells) includes wasteful ranges, such as $[16,23]$.
Therefore, we seek to implement a pluggable component to divide a z-ordered range into refined z-ordered subranges.
In the case of \Fig{\ref{fig:z-curve}}, if $[12,51]$ is divided into $[12,15]$, $[24,27]$, $[36,39]$, and $[48,51]$ before querying, then this multi-dimensional range query can be performed with no wasteful ranges.

Although such a method for refining z-ordered ranges has been proposed previously~\cite{url:dynamodb}, it was computationally expensive.
Amazon DynamoDB teams introduced a \emph{jump-in} method for refining z-ordered ranges that computes the next starting point of a divided z-ordered range~\cite{appinfo:Tropf1981,vldb:Ramsak2000}.
Their approach can divide a z-ordered range correctly, but its computational cost increases exponentially with the number of dimensions.
That is, if high-dimensional data are used, then the computational time to refine a z-ordered range becomes a bottleneck.

In this paper, we propose a \emph{jump-out} method and approximating z-ordered query regions to achieve efficient range refinement.
We summarize our contributions as follows.
\begin{itemize}
  \item We introduce a jump-out method that computes the next endpoint of a divided z-ordered range.
        Using both the jump-in and jump-out methods reduces the computational cost of refining a z-ordered range, and a z-ordered range cannot be approximated without the jump-out method.
  \item We propose approximating z-ordered query regions.
        Even if both the jump-in and jump-out methods are used, the computational time required to divide a z-ordered range is expensive.
        Therefore, we allow some wasteful regions while keeping the multi-dimensional locality, thereby reducing drastically the time taken to refine a z-ordered range.
  \item The experimental results demonstrate that our approach can divide a z-ordered range within a reasonable execution time.
        Thus, the proposed method can be used as pluggable preprocessing for multi-dimensional range queries in the z-order curve space.
        To show an example, we implemented the proposed method using PL/pgSQL in PostgreSQL and compared fundamental indices for multi-dimensional range queries.
\end{itemize}

The rest of this paper is organized as follows.
We present related work in \Sec{\ref{sec:related-work}}, and in \Sec{\ref{sec:preliminaries}} we introduce fundamental concepts needed to discuss the proposed method.
We introduce the jump-out algorithm in \Sec{\ref{sec:jump-out}} and the approximation of a z-ordered query region in \Sec{\ref{sec:approximation}}.
We evaluate the proposed method experimentally in \Sec{\ref{sec:experiments}}, and we conclude the paper in \Sec{\ref{sec:conclusion}}.

\section{Related Work}\label{sec:related-work}

Some previous studies used the z-order curve space to construct their systems~\cite{tech:Bayer1998,vldb:Ramsak2000,dpdb:nishimura2013,ieice:Le2018,vldb:Armbrust2020}.
In particular, UB-tree (Universal B-tree)~\cite{tech:Bayer1998,vldb:Ramsak2000} is an extension of B-tree and retains one z-ordered range in one leaf page; that is, UB-tree divides the z-order curve space into subregions based on data density.
If leaf pages are entirely in wasteful regions, the UB-tree skips them to improve the performance of range queries.
Because UB-tree avoids fetching such useless pages by using the jump-in method~\cite{vldb:Ramsak2000}, it achieves efficient multi-dimensional range queries in the z-order curve space.
$\mathcal{MD}$-HBase~\cite{dpdb:nishimura2013} uses a similar approach to that of UB-tree in the environment of distributed database systems (Apache HBase~\cite{url:hbase}).
However, unlike UB-tree, $\mathcal{MD}$-HBase uses spatial indices ($kd$-tree and quad-tree) as the top layer of its index and divides the z-order curve space.
$\mathcal{MD}$-HBase summarizes partitioned regions by using a common prefix of z-ordered values, and it uses such prefixes to prune useless regions in multi-dimensional query processing.
G-HBase~\cite{ieice:Le2018} uses the z-order curve space as Geohash to process geographical data, and Delta Lake~\cite{vldb:Armbrust2020} also uses it to perform multi-dimensional range queries in columnar formats.

Unlike previous studies, we aim to use existing one-dimensional indices.
The previous studies implemented their indices by using the z-order curve space and additional techniques, but in doing so they missed the opportunity to use novel one-dimensional indices.
State-of-the-art indices such as Masstree~\cite{eurosys:Mao2012} and P-tree~\cite{vldb:Sum2019} were developed for the recent in-memory and many-core environment.
Because new databases will implement these indices for the current environment, we consider using them for efficient multi-dimensional range queries.
We can also improve the performance of existing methods such as $\mathcal{MD}$-HBase by integrating them with the proposed method.

\section{Preliminaries}\label{sec:preliminaries}

In this section, we describe basic concepts needed to discuss the proposed method.
We begin by introducing our target data and their conversion into the z-order curve space.
We then define multi-dimensional range queries and z-ordered range queries.
Finally, we explain the existing jump-in algorithm.

\subsection{Data Definition}

In this paper, we use multi-dimensional data as our target data $\Table$.
A table $\Table$ has multiple columns $\Columns$, and $\Columns$ is divided into three parts: (i) a primary key $\Column^{pk}$, (ii) selection columns $\Columns^{s} = \{ \Column^{s}_{1}, \ldots, \Column^{s}_{m} \}$, and (iii) value columns $\Columns^{v}$.
A primary key $\Column^{pk}$ uniquely identifies each tuple in $\Table$, but we assume that it has no semantics (i.e., a surrogate key) and is not used to select tuples.
Instead of a primary key, we use selection columns $\Columns^{s}$ to select tuples.
We assume that selection columns are correlated with each other (e.g., spatiotemporal data), and therefore multi-dimensional locality is important.
In the following, we use multi-dimensional data and selection columns interchangeably.
Value columns $\Columns^{v}$ are optional and represent specific values of each tuple.

The z-order curve~\cite{tech:Morton1966} is a space-filling curve and converts multi-dimensional data into one-dimensional data.
An example of a two-dimensional case is shown in \Fig{\ref{fig:z-curve}}, in which each selection column is represented as discrete attributes and each cell contains a bit string and its decimal value.
Intuitively, a z-order curve converts multi-dimensional data by arranging bits alternately.
In a two-dimensional case, odd-numbered bits represent the first dimension $\Column^{s}_{1}$, and even-numbered bits represent the second dimension $\Column^{s}_{2}$.
As a result, z-ordered values draw a ``Z'' curve in a two-dimensional space, such as in the range $[0,3]$.
We use these z-ordered values as an additional selection column $\Column^{z}$ to process multi-dimensional range queries efficiently.

In summary, our target table $\Table$ comprises columns as follows:
\begin{equation}
  \Table \equiv \{ \Column^{pk} \} \cup \Columns^{s} \cup \{ \Column^{z} \} \cup \Columns^{v}.
\end{equation}

\subsection{Multi-dimensional and Z-ordered Range Queries}

Multi-dimensional range queries have range conditions for each dimension to select tuples, thereby causing an overall query region to become a hyper-rectangle in a multi-dimensional space.
In the two-dimensional case of \Fig{\ref{fig:z-curve}}, the query region is a rectangle that is represented as the intersection of range conditions $\Value^{s}_{1} \in [2,5]$ and $\Value^{s}_{2} \in [2,5]$.

Because the z-order curve has a simple structure, a multi-dimensional range query is easily converted into a z-ordered range query.
Z-ordered values have the property
\begin{equation}
  \forall \mathbf{x}, \mathbf{y} \in \Columns^{s}, \bigwedge_{\Column^{s}_{i} \in \Columns^{s}} x_{i} \leq y_{i} \rightarrow f_{z}(\mathbf{x}) \leq f_{z}(\mathbf{y}), \label{eq:z-order-property}
\end{equation}
where the function $f_{z}: \Columns^{s} \rightarrow \Column^{z}$ converts a multi-dimensional value into a z-ordered one.
When we convert the smallest and largest coordinates in a query region into z-ordered values, all the coordinates in the query region must be within its z-ordered range.
In the case of \Fig{\ref{fig:z-curve}}, the z-ordered range of the smallest coordinate $(2,2) \in \Column^{s}_{1} \times \Column^{s}_{2}$ and the largest coordinate $(5,5) \in \Column^{s}_{1} \times \Column^{s}_{2}$ is $[12,51]$ and includes all the coordinates in the query region, namely, $[12,15], [24,27], [36,39]$, and $[48,51]$.

As introduced in \Sec{\ref{sec:introduction}}, a z-ordered range does not correspond exactly to the original multi-dimensional range.
In the previous example, the original two-dimensional query region is shown by the solid rectangle, but the z-ordered range spreads out as shown by the gray cells.
Therefore, the ranges $[16,23], [28,35]$, and $[40,47]$ are wasteful for answering the query, and such wasteful ranges must be pruned if z-ordered range queries are to be answered efficiently.

\subsection{Z-ordered Range Refinement with the Jump-in Algorithm}

The previous study~\cite{url:dynamodb} introduced dividing a z-ordered range by using a jump-in algorithm~\cite{appinfo:Tropf1981,vldb:Ramsak2000}, for which the BIGMIN algorithm~\cite{appinfo:Tropf1981} and the getNextZValue algorithm~\cite{vldb:Ramsak2000} were proposed.
Given a z-ordered query range and a z-value outside the original query region, the jump-in algorithm computes the next z-ordered value that intersects the original query region.
For example, in \Fig{\ref{fig:z-curve}}, if we input $28$ as the current z-ordered value and $[12,51]$ as the z-ordered query range, the jump-in algorithm returns $36$ as the next intersecting z-ordered value.
Previous studies divided the original z-ordered range by using this jump-in algorithm, as shown in \Fig{\ref{fig:algo:jump-in}}.
The jump-in-based range refinement checks whether each z-ordered position is in the original query region.
If the current position is outside the query region, then the algorithm used the BIGMIN or getNextZValue algorithm to obtain the next intersecting position.

\begin{algorithm}[t]
  \small
  \DontPrintSemicolon
  \SetKwProg{Proc}{Function}{}{}
  \SetKwFunction{JumpIn}{getNextJumpIn}
  \KwIn{$[\ZValue_{s}, \ZValue_{e}]$ \tcp*{a z-ordered range}}
  $\Ranges^{z} \leftarrow \emptyset$ \tcp*{refined z-ordered ranges}
  $\ZValue_{s}' \leftarrow \ZValue_{s}$ \tcp*{a starting position of a divided range}
  $\ZValue_{c} \leftarrow \ZValue_{s}'$ \tcp*{a current z-ordered position}
  \While{$\ZValue_{c} < \ZValue_{e}$}{
    \uIf{$\ZValue_{c}$ is in an original query region}{
      $\ZValue_{c} \leftarrow \ZValue_{c} + 1$
    }
    \Else{
      $\Ranges^{z} \leftarrow \Ranges^{z} \cup \{ [\ZValue_{s}', \ZValue_{c} - 1] \}$\;
      $\ZValue_{s}' \leftarrow \JumpIn(\ZValue_{c}, [\ZValue_{s}, \ZValue_{e}])$ \tcp*{BIGMIN or getNextZValue}
      $\ZValue_{c} \leftarrow \ZValue_{s}'$
    }
  }
  $\Ranges^{z} \leftarrow \Ranges^{z} \cup \{ [\ZValue_{s}', \ZValue_{e}] \}$\;
  \Return{$\Ranges^{z}$}
  \caption{Jump-in-based algorithm for range refinement}
  \label{fig:algo:jump-in}
\end{algorithm}

Although the jump-in-based range refinement can divide a z-ordered range into fragments of exact query regions, doing so can take considerable computation time.
The jump-in-based method checks whether the current position is in the original query region (lines 5 and 6), but the number of such checks increases exponentially with the number of dimensions.
If dealing with $m$-dimensional data $\Columns^{s} = \{ \Column^{s}_{1}, \ldots, \Column^{s}_{m} \}$, the algorithm must check
\begin{equation}
  Volume_{m} = \prod_{i = 1}^{m} (e_{i} - s_{i} + 1)
\end{equation}
z-ordered values, where $s_{i}, e_{i} \in \Column^{s}_{i}$ are the start and end values, respectively, of a query range for each dimension $\Column^{s}_{i}$; that is, the entire interior of a hyper-rectangle must be scanned.
Because the volume of a hyper-rectangle increases drastically with high-dimensional data, a more efficient approach is required.

\section{Range Refinement with the Jump-out Algorithm}\label{sec:jump-out}

In this section, we introduce a jump-out algorithm to divide a z-ordered range efficiently.
Intuitively, given a current z-ordered position, the jump-out algorithm computes the next-smallest z-ordered value outside the original query region.
In the case of \Fig{\ref{fig:z-curve}}, if we give current position $12$ and z-ordered query range $[12,51]$ to the jump-out algorithm, it returns $16$ as the next z-ordered value.
Thus, we can generate a refined fragment $[12,15]$ from this result without scanning the inside of a hyper-rectangle.

To discuss the jump-out algorithm, we define the \emph{layers} of a z-curve.
If we use $l$ bits to express the values for each dimension in $\Columns^{s} = \{ \Column^{s}_{1}, \ldots, \Column^{s}_{m} \}$, then we consider a z-curve to have $m \cdot l$ layers.
For example, \Fig{\ref{fig:z-curve}} shows a z-curve that has $2 \cdot 3 = 6$ layers.
In other words, the number of layers is equivalent to the number of bits required to express z-ordered values.

\begin{figure}[t]
  \includegraphics{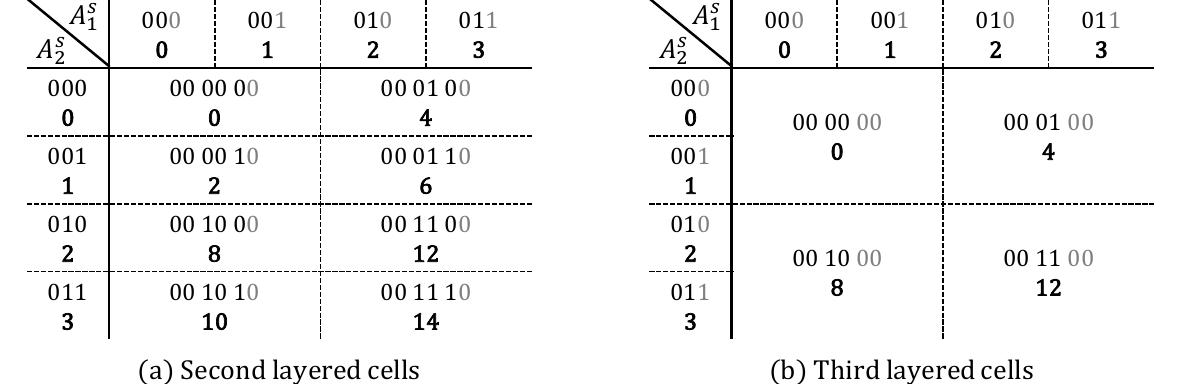}
  \caption{Layered cells for second and third layers in z-order curve space}

  \label{fig:layered-cells}
\end{figure}

We also consider \emph{layered cells} for each layer, which express the smallest units therein.
For example, \Fig{\ref{fig:layered-cells}} shows layered cells in the second and third layers.
Note that layered cells in the first layer correspond to the original z-order curve space.
In other words, layered cells are z-order curve space modified by filling lower bits with zeros.
In the case of \Fig{\ref{fig:layered-cells}}, the first bit in the second layer and the first and second bits in the third layer are filled with zeros.

In the following, we explain the jump-out algorithm and then describe a method for refining a z-ordered range using both the jump-in and jump-out algorithms.

\subsection{Jump-out Algorithm}

Figure~\ref{fig:algo:jump-out} shows our jump-out algorithm.
Note that we assume that an input z-ordered value is provided by the jump-in algorithm.
Our jump-out algorithm has two phases: forward sliding out (lines~1--5) and backward carrying out (lines~6--14).
In the following, we explain each phase in detail.

\begin{algorithm}[t]
  \small
  \DontPrintSemicolon
  \SetKwProg{Proc}{Function}{}{}
  \SetKwFunction{JumpIn}{getNextJumpIn}
  \SetKwFunction{JumpOut}{getNextJumpOut}
  \KwIn{$\ZValue_{c}, [\ZValue_{s}, \ZValue_{e}]$ \tcp*{a current z-order value and a z-ordered range}}
  $\ZValue_{min} \leftarrow \ZValue_{e}$\;
  \For{$i = 1$ \KwTo $m$}{
    $\ZValue_{out} \leftarrow$ slide $\ZValue_{c}$ outside a query region over dimension $i$\;
    \If{$\ZValue_{out} < \ZValue_{min}$}{
      $\ZValue_{min} \leftarrow \ZValue_{out}$
    }
  }
  \For{$i = 2$ \KwTo $m \cdot l$}{
    \If{bit $i$ of $\ZValue_{c}$ is zero}{
      \textbf{continue}
    }
    $\ZValue_{out} \leftarrow$ move $\ZValue_{c}$ to the next $i$-th layered cell\;
    \uIf{$\ZValue_{out} \geq \ZValue_{min}$}{
      \textbf{break}
    }\ElseIf{$\ZValue_{out}$ is outside a query region}{
      $\ZValue_{min} \leftarrow \ZValue_{out}$\;
      \textbf{break}
    }
  }
  \Return{$\ZValue_{min}$}
  \caption{Jump-out algorithm}
  \label{fig:algo:jump-out}
\end{algorithm}

\begin{figure}[t]
  \includegraphics{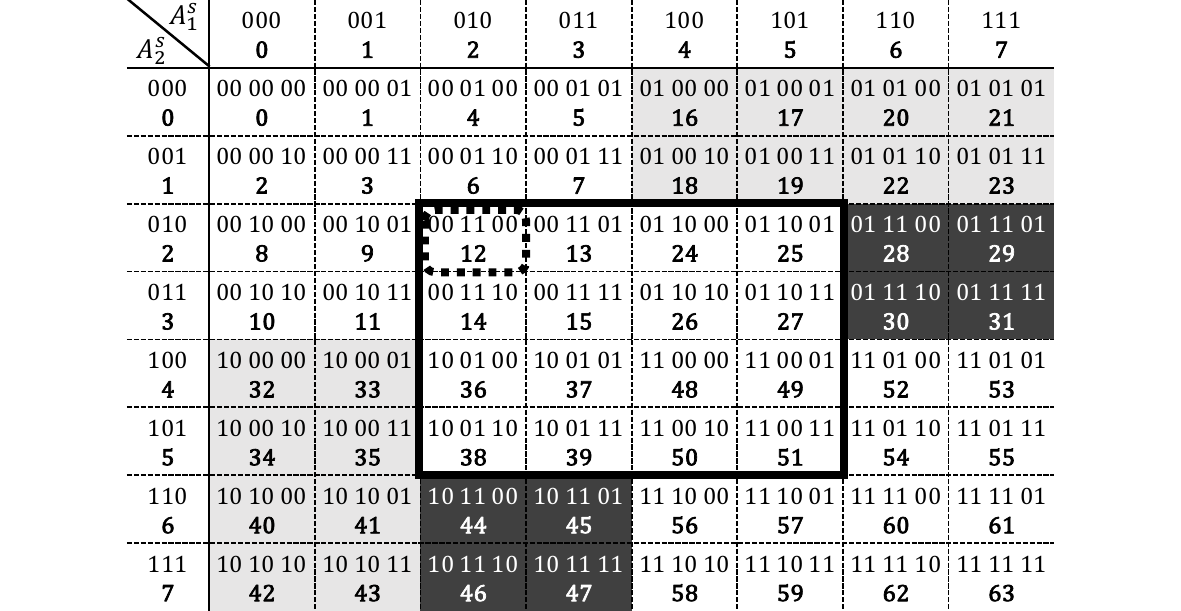}
  \caption{Jump-out candidate regions partitioned by \Eq{\ref{eq:z-order-property}}: the regions following \Eq{\ref{eq:z-order-property}} are colored dark gray, and the remaining ones are colored light gray}
  \label{fig:jump-out-candidates}
\end{figure}

First, we slide the current z-ordered position forward in each dimension to obtain an initial jump-out candidate.
Given a candidate jump-out region, we divide it into two regions as shown in \Fig{\ref{fig:jump-out-candidates}}; that is, the region follows \Eq{\ref{eq:z-order-property}} (dark-gray cells) and the remaining region (light-gray cells).
Note that we use $12$ as the current z-ordered value in \Fig{\ref{fig:jump-out-candidates}}.
When a candidate region follows \Eq{\ref{eq:z-order-property}}, the smallest value therein is any of the results of sliding out over each dimension.
In the two-dimensional case, the jump-out candidates are $28$ and $44$ in \Fig{\ref{fig:jump-out-candidates}} because the smallest value is any of the left-upper positions in each region.
Thus, we select $28$ as the initial jump-out candidate in this case.

Next, we move the current z-ordered position backward in any of the original dimensions to check the remaining candidate regions.
Because the remaining candidate regions (light-gray cells in \Fig{\ref{fig:jump-out-candidates}}) do not follow \Eq{\ref{eq:z-order-property}}, we require backward movements in any of the original dimensions.
In the case of \Fig{\ref{fig:jump-out-candidates}}, we require a backward movement in dimension $\Column^{s}_{2}$ to move to $16$ (from $2$ to $0$ in dimension $\Column^{s}_{2}$).
To achieve such backward movements, we must carry out an upper bit.
In \Fig{\ref{fig:jump-out-candidates}}, because the current z-ordered value $12$ has the bit string \texttt{001100}, we must carry out the sixth bit to reset the fifth bit, such as \texttt{010000} ($16$ in decimal).

\begin{figure}[t]
  \includegraphics{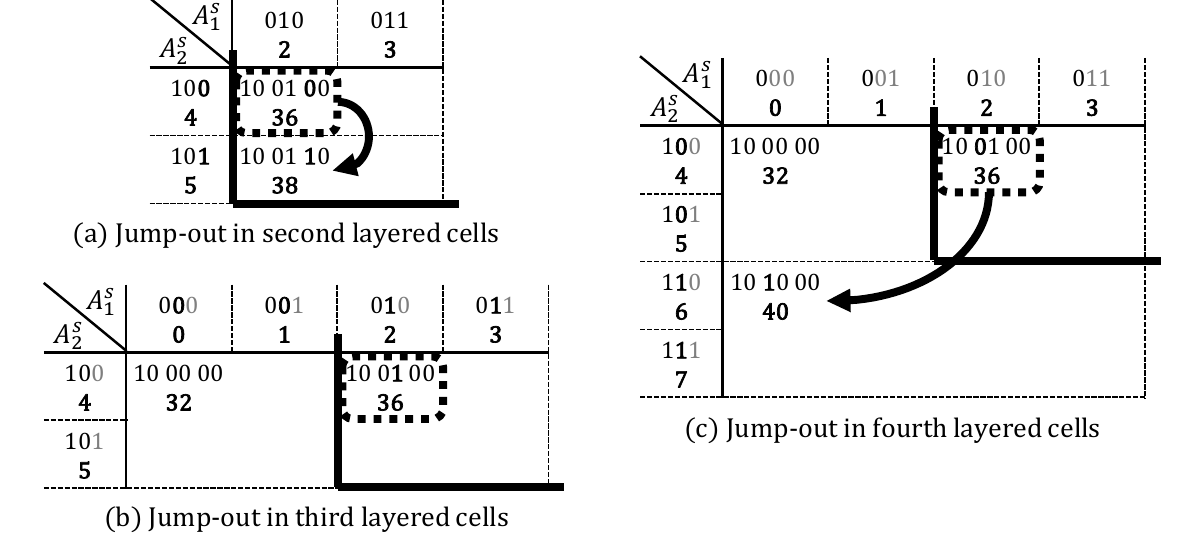}
  \caption{Backward movements based on layered cells for current z-ordered value $36$ and z-ordered query region $[12,51]$}

  \label{fig:backward-example}
\end{figure}

In our jump-out algorithm, we use layered cells to carry out an upper bit.
The jump-out algorithm checks each bit $i$ of a current z-ordered value and tries to jump to the next $i$-th layered cell if bit $i$ is zero.
In other words, the algorithm tries to perform a backward movement in layer $i-1$ by carrying out bit $i$.
Figure~\ref{fig:backward-example} shows an example involving current z-ordered value $36$ and z-ordered query range $[12,51]$.
In the second layer [\Fig{\ref{fig:backward-example}}(a)], because the second bit of current z-ordered value $36$ (\texttt{1001\textit{0}0}) is zero, the algorithm tries to move backward in the first layer (from \texttt{011} to \texttt{010} in dimension $\Column^{s}_{1}$).
However, we continue the jump-out algorithm because the next second-layered cell $38$ is in the original query region.
In the third layer [\Fig{\ref{fig:backward-example}}(b)], because the third bit of $36$ (\texttt{100\textit{1}00}) is one, the algorithm skips this layer.
In the fourth layer [\Fig{\ref{fig:backward-example}}(c)], because the fourth bit of $36$ (\texttt{10\textit{0}100}) is zero, the algorithm tries to move backward in the third layer (from \texttt{010} to \texttt{000} in dimension $\Column^{s}_{1}$).
Because the next fourth-layered cell $40$ is outside the original query region, the algorithm returns $40$ as the next jump-out value.

\subsection{Z-ordered Range Refinement with the Jump-in and Jump-out Algorithms}\label{sec:jump-out:jump-in-out}

We can simplify the algorithm for z-ordered range refinement by using both the jump-in and jump-out methods.
Figure~\ref{fig:algo:jump-inout} shows our algorithm for z-ordered range refinement, which jumps in and out of the query region repeatedly, thereby avoiding having to scan the entire interior of a hyper-rectangle.

\begin{algorithm}[t]
  \small
  \DontPrintSemicolon
  \SetKwProg{Proc}{Function}{}{}
  \SetKwFunction{JumpIn}{getNextJumpIn}
  \SetKwFunction{JumpOut}{getNextJumpOut}
  \KwIn{$[\ZValue_{s}, \ZValue_{e}]$ \tcp*{a z-ordered range}}
  $\Ranges^{z} \leftarrow \emptyset$ \tcp*{refined z-ordered ranges}
  $\ZValue_{s}' \leftarrow \ZValue_{s}$ \tcp*{a starting position of a divided range}
  $\ZValue_{e}' \leftarrow \JumpOut(\ZValue_{s}', [\ZValue_{s}, \ZValue_{e}])$\;
  \While{$\ZValue_{e}' < \ZValue_{e}$}{
    $\Ranges^{z} \leftarrow \Ranges^{z} \cup \{ [\ZValue_{s}', \ZValue_{e}' - 1] \}$\;
    $\ZValue_{s}' \leftarrow \JumpIn(\ZValue_{e}', [\ZValue_{s}, \ZValue_{e}])$\;
    $\ZValue_{e}' \leftarrow \JumpOut(\ZValue_{s}', [\ZValue_{s}, \ZValue_{e}])$\;
  }
  $\Ranges^{z} \leftarrow \Ranges^{z} \cup \{ [\ZValue_{s}', \ZValue_{e}] \}$\;
  \Return{$\Ranges^{z}$}
  \caption{Algorithm for range refinement with the jump-in and jump-out methods}
  \label{fig:algo:jump-inout}
\end{algorithm}

However, the range refinement still involves non-negligible computational complexity.
Although our algorithm uses the jump-out method to avoid having to scan the inside of a hyper-rectangle, we must jump in and out over its surface.
That is, given $m$-dimensional data $\Columns^{s} = \{ \Column^{s}_{1}, \ldots, \Column^{s}_{m} \}$, the algorithm requires
\begin{equation}
  Surface_{m} = 2 \cdot \sum_{i = 1}^{m} \prod_{j = 1 (j \neq i)}^{m} (e_{j} - s_{j} + 1)
\end{equation}
jumps, where $s_{i}, e_{i} \in \Column^{s}_{i}$ are the start and end values, respectively, of a query range for each dimension $\Column^{s}_{i}$.
Therefore, we must construct a more efficient method for range refinement before query execution.

\section{Approximation of Z-ordered Range Queries}\label{sec:approximation}

Because our algorithm for range refinement in \Fig{\ref{fig:algo:jump-inout}} is still slow, we propose an approximation of a z-ordered range.
As described in \Sec{\ref{sec:jump-out:jump-in-out}}, many jumps over the surface of a hyper-rectangle are required to refine a z-ordered range; therefore we reduce the computational cost of doing so by permitting some wasteful ranges.

\begin{figure}[t]
  \includegraphics{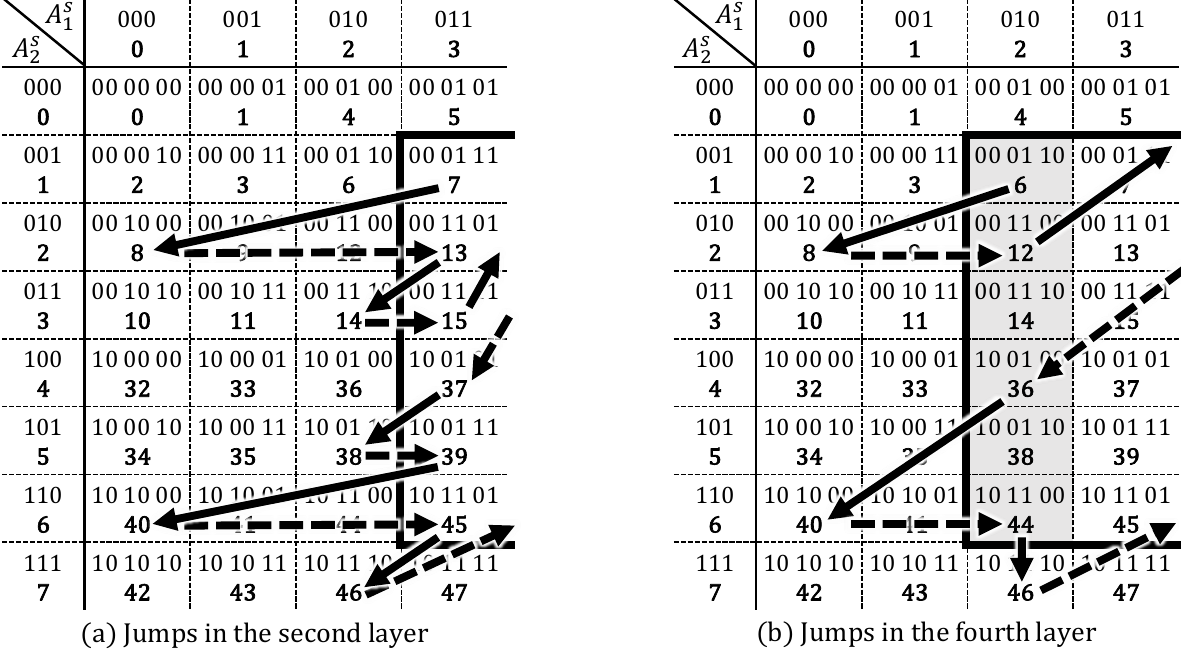}
  \caption{Reducing the number of jumps based on z-ordered range approximation}
  \label{fig:z-range-approximation}
\end{figure}

An important point to note when seeking to reduce the computational cost is that the jumping in and out is processed based on layered cells.
That is, the jump-in/out is larger if a query region touches the borders of layered cells in the upper layers.
As an example, consider the z-ordered query ranges $[7,57]$ and $[6,57]$.
As shown in \Fig{\ref{fig:z-range-approximation}}, range $[7,57]$ touches the borders of layered cells in the first layer (\texttt{011} in dimension $\Column^{s}_{1}$), thus the algorithm executes jump-out for each second-layered cell, such as the jump from $13$ to $14$.
Meanwhile, because range $[6,57]$ touches the borders of layered cells in the third layer (\texttt{010} in dimension $\Column^{s}_{1}$), the jump-out is executed in the fourth layer.
Consequently, the jump-out skips four cells in \Fig{\ref{fig:z-range-approximation}}(b), such as the jump from $36$ to $40$.

Thus, we expand a z-ordered range based on layers to reduce the number of jumps.
Although the expansion also generates wasteful ranges, multi-dimensional locality is retained.
In the case of \Fig{\ref{fig:z-range-approximation}}, if we consider range $[7,57]$ as the original z-ordered range, then the expanded range $[6,57]$ includes wasteful ones such as $[6,6]$.
However, unlike the case of \Fig{\ref{fig:z-curve}}, we can control wasteful regions by using multi-dimensional ranges.

To approximate a query region, we compute the surface areas on a hyper-rectangle and then expand the range that has the largest area.
We calculate each surface area on dimension $\Column^{s}_{i}$ roughly by using
\begin{align}
  Surface_{i}^{s} & = \frac{\prod_{j = 1 (j \neq i)}^{m} (e_{j} - s_{j} + 1)}{2^{(m-1)b^{s} + i - 1}} \quad \textrm{and} \\
  Surface_{i}^{e} & = \frac{\prod_{j = 1 (j \neq i)}^{m} (e_{j} - s_{j} + 1)}{2^{(m-1)b^{e} + i - 1}},
\end{align}
where $b^{s}$ is the number of continuous zero bits in the postfix of $s_{i}$, and $b^{e}$ is the number of continuous one bits in the postfix of $e_{i}$.
That is, we calculate each surface area of a query region based on corresponding layed cells.
We then select the largest surface area and expand the corresponding range by filling $s_{i}$ or $e_{i}$ with a zero or one bit.
For example, we calculate $Surface_{1}^{s}$ in \Fig{\ref{fig:z-range-approximation}}(a) as
\begin{equation}
  Surface_{1}^{s} = \frac{\prod_{j = 1 (j \neq i)}^{2} (e_{j} - s_{j} + 1)}{2^{0}} = \frac{6 - 1 + 1}{1} = 6.
\end{equation}
Because this surface area is the largest one, we expand $s_{1}$ to $2$ (\texttt{010}) from $3$ (\texttt{011}) and update its estimated surface area as
\begin{equation}
  Surface_{1}^{s} = \frac{\prod_{j = 1 (j \neq i)}^{2} (e_{j} - s_{j} + 1)}{2^{1}} = \frac{6 - 1 + 1}{2} = 3.
\end{equation}
Although estimated surface areas are different from the exact number of jumps, we use them as a simple cost function.
We continue this process until the total surface area on a hyper-rectangle becomes sufficiently small.

\section{Experiments}\label{sec:experiments}

In this section, we evaluate the proposed method on PostgreSQL by using synthetic datasets.

\paragraph{Implementation}

We implemented the proposed method by using PL/pgSQL on PostgreSQL~\cite{url:artifact}; for the proposed method, we wrote user-defined functions and we neither modified the PostgreSQL source code nor extended functionalities such as types and indices.
To approximate z-ordered range queries, we used $500$ as the threshold for the estimated total surface areas of a query region; note that we determined this threshold empirically because it depends on the execution environment.
Table~1 summarizes the experimental environment.

\begin{table}[t]
  \caption{Experimental environment}
  \label{tab:environment}
  \setlength{\tabcolsep}{2truemm}
  \centering \small
  \begin{tabular}{ll}
    \toprule
    Item       & Value                                     \\
    \cmidrule(r){1-1}\cmidrule(l){2-2}
    CPU        & Intel(R) Xeon(R) Gold 6262V (two sockets) \\
    RAM        & 219~GB                                    \\
    OS         & Ubuntu 18.04.5 LTS                        \\
    PostgreSQL & 12.4-1.pgdg18.04+1                        \\
    PostGIS    & 3.0.2+dfsg-4.pgdg18.04+1                  \\
    \bottomrule
  \end{tabular}
\end{table}

\paragraph{Datasets}

We used synthetic datasets to evaluate our approach.
We generated five datasets with 10 million tuples, and each dataset had corresponding selection columns $\Columns^{s}$ from two to six dimensions.
The value range of each selection column was $[-512,512]$, and we assigned values randomly for each tuple.
Each dataset had a z-ordered column $\Column^{z}$ but no value columns $\Columns^{v}$.
Note that we used 10 bits to express each selection column $\Column^{s}_{i}$ in the z-order curve space.
Each two- or three-dimensional dataset also had a PostGIS geometry column so that we could compare our approach with that of spatial indices.

\paragraph{Comparison Methods}

We compare the proposed method with fundamental indices in PostgreSQL and PostGIS to show that our approach achieves sufficiently effective performance.
In the experiments, we used four indices: independent indices, a multi-column index, and GiST/SPGiST indices~\cite{vldb:Hellerstein1995,spgist2001} on PostGIS geometries.
Because GiST/SPGiST indices work as spatial indices (such as R-tree and $kd$-tree) on PostGIS geometries, we use them as two-/three-dimensional spatial indices.

\subsection{Efficiency of Z-ordered Range Refinement}

\begin{figure}[t]
  \includegraphics{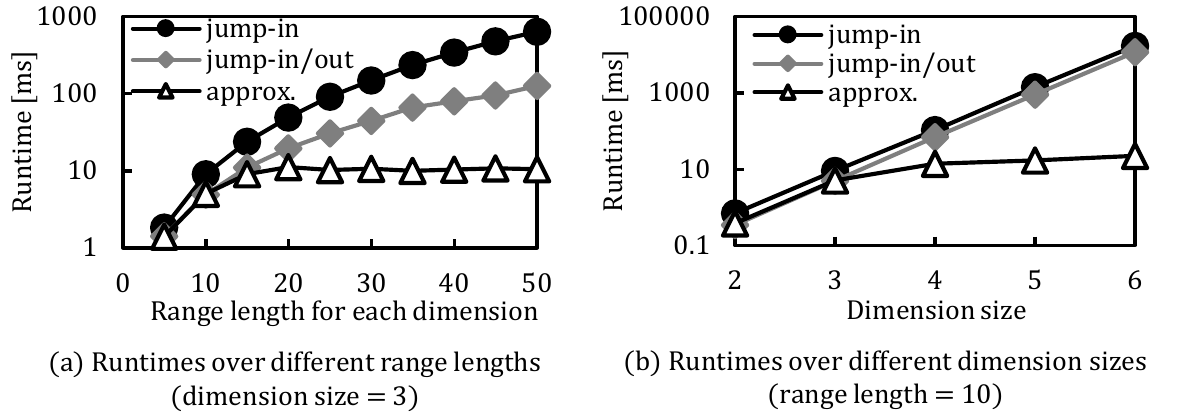}
  \caption{Runtimes of z-ordered range refinement}
  \label{fig:runtime-range-refinement}
\end{figure}

First, we evaluate the efficiency of the proposed z-ordered range refinement.
Figure~\ref{fig:runtime-range-refinement} shows the runtimes of z-ordered range refinement for each method.
In the experiments, we changed the lengths of the ranges for each dimension and selected query regions randomly.
Figure~\ref{fig:runtime-range-refinement} shows the average of 100 measurements.
Note that we show only one case for each horizontal axis (range length and dimension size) because the other graphs have similar results.

When we use only the jump-in and jump-out methods, the runtimes increase rapidly over both range length and dimension size.
Although we can moderate the runtime increase rate by using the jump-out method, the runtimes of jump-in/out are still long.
In particular, the runtimes of jump-in/out increase exponentially with the dimension size.
These results indicate that the simple z-ordered range refinement is inefficient as preprocessing for z-ordered range queries.

By contrast, the proposed approach achieves efficient z-ordered range refinement.
Because we approximate a z-ordered range, the proposed method limits the number of jumps over the surface of a query region.
Consequently, the upper runtime limit is roughly 10~ms in this implementation, which is sufficiently rapid execution to use as preprocessing for z-ordered range queries.

\begin{figure}[t]
  \includegraphics{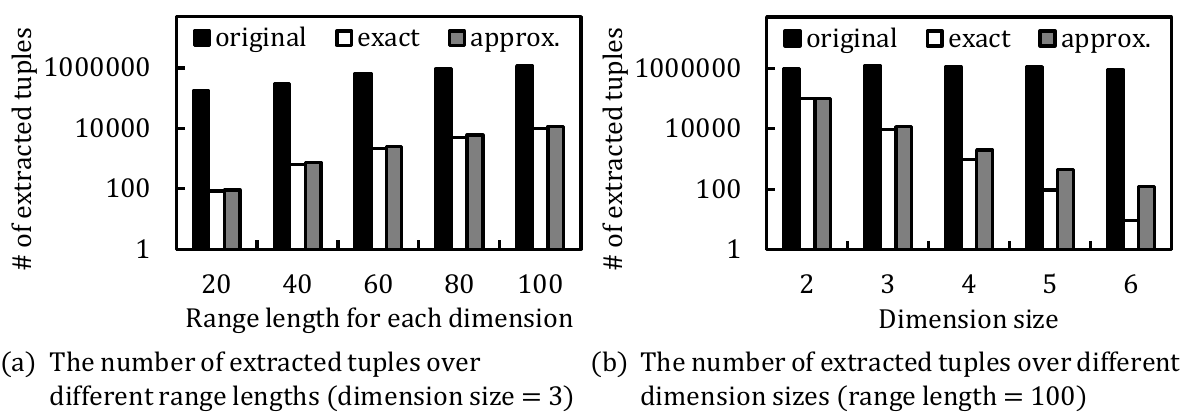}
  \caption{The number of extracted tuples with/without range refinement}
  \label{fig:range-reduction}
\end{figure}

Furthermore, the approximation of a z-ordered range has only a limited effect on the amount of extracted tuples.
Figure~\ref{fig:range-reduction} shows the number of extracted tuples of z-ordered range queries.
When we use the original z-order range directly, we extract a lot of unnecessary tuples due to wasteful ranges.
In comparison with the original z-ordered range, refined z-ordered ranges (\texttt{exact/approx.} in \Fig{\ref{fig:range-reduction}}) can reduce the number of tuples significantly.
Although approximated z-ordered ranges have more unnecessary tuples as compared with exact ones, it sufficiently suppresses the number of extracted tuples.

\subsection{Effectiveness of Pluggable Z-ordered Range Refinement}

\begin{figure}[t]
  \includegraphics{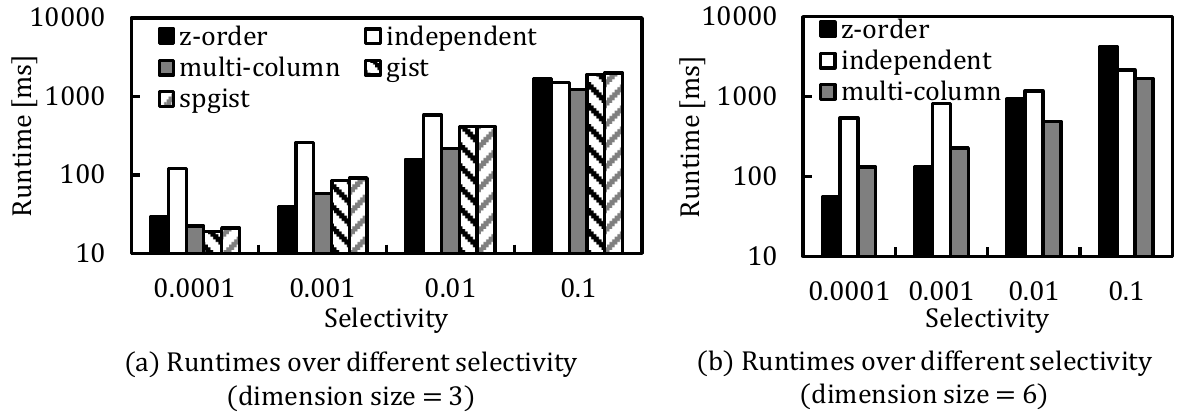}
  \caption{Runtimes of multi-dimensional range queries}

  \label{fig:runtime-range-queries}
\end{figure}

In this subsection, we demonstrate that our approach achieves sufficient performance for multi-dimensional range queries.
Figure~\ref{fig:runtime-range-queries} shows the runtimes of multi-dimensional range queries for each method over different selectivity.
In the experiments, we used the same range lengths for each dimension according to selectivity, and then we selected query regions randomly.
So that the fetched data had uniform size, we projected only a primary key $\Column^{pk}$ from tuples.
Figure~\ref{fig:runtime-range-queries} shows the average of 100 measurements.
Note that we show only the experimental results for three and six dimensions because the others are similar.

The experimental results demonstrate that the performance of our approach is either more efficient than or at least comparable to that with the existing indices without implementing a new index.
On the one hand, for sufficiently low selectivity, the runtimes of the proposed method are less than those of the others; that is, although the z-ordered range refinement adds an additional overhead, it is negligible for executing multi-dimensional range queries.
On the other hand, for high selectivity (e.g., $0.1$), the proposed method is slower than the others.
The main reason for this degraded performance is the z-ordered range approximation; because a query region becomes a large hyper-rectangle with high selectivity, the approximated z-ordered range includes more wasteful query regions and degrades the query performance.

\begin{figure}[t]
  \includegraphics{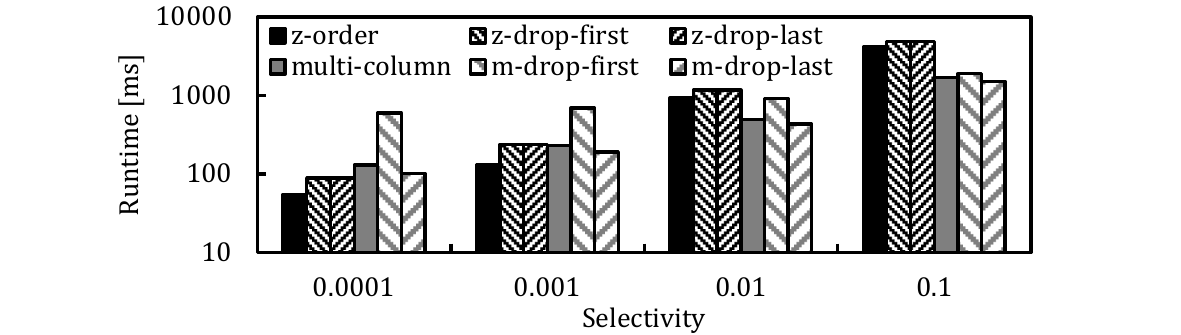}
  \caption{Runtimes of partial six-dimensional range queries: \texttt{drop-first/last} means that the first/sixth dimension is removed from the condition of range queries}

  \label{fig:runtime-partial-range-queries}
\end{figure}

We also measured the runtimes of partial multi-dimensional range queries to compare with multi-column indices.
In that experiment, we dropped one dimension from the multi-dimensional range queries.
Figure~\ref{fig:runtime-partial-range-queries} shows the runtimes of these range queries when using the proposed method and multi-column indices.
Because the effectiveness of the multi-column indices depends on the order of the columns, the query performance also deteriorates according to the dropped dimension.
Thus, dropping the first dimension from multi-dimensional range queries would increase considerably the runtimes based on multi-column indices; however, because z-ordered range queries have no such limitation, the proposed method achieves stable query performance.

\section{Conclusion}\label{sec:conclusion}

In this paper, we proposed a method for z-ordered range refinement to use existing one-dimensional indices.
We introduced the jump-out algorithm to reduce the computational cost of z-ordered range refinement, and then we proposed the approximation of a z-ordered range.
The experimental results demonstrated that we can use the proposed z-ordered range refinement as preprocessing for multi-dimensional range queries.
Furthermore, we compared our approach with the fundamental indices for multi-dimensional range queries and measured the performance of the proposed method as being either better or at least comparable.

In future work, we intend to subject the proposed method to exhaustive experiments and parallelization.
Because the results of z-ordered range refinement are exclusive z-ordered ranges, we can safely parallelize read operations for them.
We will also apply the proposed method to key-value stores (e.g., RocksDB) and evaluate the scalability of parallel z-ordered range queries.

\subsection*{Acknowledgements}

This work is based on results obtained from a project, JPNP16007, commissioned by the New Energy and Industrial Technology Development Organization (NEDO).
In addition, this work was supported partly by KAKENHI (JP20K19804, JP21H03555, and JP22H03594).

\bibliographystyle{splncs04}
\bibliography{references}

\end{document}